\documentstyle[12pt,epsfig,axodraw]{article}

\newcommand{\ee}{e^+e^-}
\newcommand{\tb}{\tan\beta}
\newcommand{\s}{\\ \vspace*{-3mm} }
\newcommand{\nn}{\noindent}

\newcommand{\mc}{m_{\tilde\chi^0_1}}

\begin{document}

\newpage
\setcounter{page}{0}

\begin{titlepage}
\begin{flushright}
\hfill{DESY 98--175}\\
\hfill{KIAS--P98038}\\
\hfill{PM/98--39}\\
\hfill{SNUTP 98--129}\\

\hfill{\today}
\end{flushright}
\vspace*{1.0cm}

\begin{center}
{\large \bf Determining SUSY Parameters \\
            in Chargino Pair--Production in $e^+e^-$ Collisions}
\end{center}
\vskip 1.cm
\begin{center}
{\sc S.Y. Choi$^1$, A. Djouadi$^2$, H.S. Song$^3$} and 
{\sc P.M. Zerwas$^4$} 

\vskip 0.8cm

\begin{small} 
$^1$ Korea Institute for Advanced Study, 207--43, Cheongryangri--dong\\
     Dongdaemun--gu, Seoul 130--012, Korea 
\vskip 0.2cm
$^2$ Phys. Math. et Th\'eorique, Universit\'{e} Montpellier II, 
     F--34095 Montpellier, France 
\vskip 0.2cm
$^3$ Center for Theoretical Physics and Department of Physics\\
     Seoul National University, Seoul 151-742, Korea
\vskip 0.2cm
$^4$  DESY, Deutsches Elektronen-Synchrotron, D--22603 Hamburg, Germany 
\end{small}
\end{center}

\vskip 2cm

\setcounter{footnote}{0}
\begin{abstract}
In most supersymmetric theories,
charginos $\tilde{\chi}^\pm_{1,2}$, mixtures of charged color--neutral 
gauginos and higgsinos, belong to the class of the lightest 
supersymmetric particles. They are easy to observe at 
$e^+e^-$ colliders. 
By measuring the total cross sections and the left--right asymmetries 
with polarized electron beams in 
$e^+e^- \rightarrow \tilde{\chi}_i^- \tilde{\chi}_j^+ \, \,  [i,j=1,2]$,
the chargino masses and the gaugino--higgsino
mixing angles can be determined. From these observables the fundamental 
SUSY parameters can be derived: 
the SU(2) gaugino mass $M_2$,
the modulus $|\mu|$ and $\cos \Phi_\mu$
of the higgsino mass parameter, and $\tan\beta 
= v_2/v_1$, the ratio of the vacuum expectation values of the two neutral
Higgs doublet fields. The solutions are unique; the CP-violating phase
$\Phi_\mu$ can be determined uniquely by analyzing effects due to
the normal polarization of the charginos.
\end{abstract}
\end{titlepage}

\newpage

\subsection*{1. Introduction}

In supersymmetric theories, the spin--1/2 partners of the $W$ bosons and 
the charged Higgs bosons, $\tilde{W}^\pm$ and $\tilde{H}^\pm$, mix to form 
chargino mass eigenstates $\tilde{\chi}^\pm_{1,2}$. The chargino mass
matrix \cite{R1} is given in the $(\tilde{W}^-,\tilde{H}^-)$ basis by 
\begin{eqnarray}
{\cal M}_C=\left(\begin{array}{cc}
                M_2                &      \sqrt{2}m_W\cos\beta  \\
             \sqrt{2}m_W\sin\beta  &             \mu   
                  \end{array}\right)
\label{eq:mass matrix}
\end{eqnarray}
which is built up by the fundamental supersymmetry (SUSY) parameters:
the SU(2) gaugino mass $M_2$, the higgsino mass parameter $\mu$, 
and the ratio $\tb=v_2/v_1$ of the vacuum expectation values of the two 
neutral Higgs fields which break the electroweak symmetry. 
In CP--noninvariant theories, the gaugino mass $M_2$ and the 
higgsino mass parameter $\mu$ can be complex. 
However, by reparametrization of the
fields, $M_2$ can be assumed real and positive without loss of
generality so that the only non--trivial invariant phase is 
attributed to $\mu$:
\begin{eqnarray}
\mu=|\mu|{\rm e}^{i\Phi_\mu}
\end{eqnarray}
The angle $\Phi_\mu$ can vary between 0 and $2\pi$.
Once charginos will have been discovered, the experimental analysis of 
their properties, production and decay mechanisms, will therefore reveal 
the basic structure of the underlying supersymmetric theory. \s

Charginos are produced in $e^+e^-$ collisions, either in diagonal or in
mixed 
pairs \cite{R2}--\cite{R2C}. In the present analysis, we will focus on all  
combinations of  chargino pairs $\tilde{\chi}^\pm_{1,2}$ 
in $e^+e^-$ collisions: 
\begin{eqnarray*}
e^+ e^- \ \rightarrow \ \tilde{\chi}^-_i \ \tilde{\chi}^+_j \ \ \ [i,j=1,2] 
\end{eqnarray*}
If the collider energy is sufficient to produce the
two  chargino states in pairs, 
the underlying fundamental SUSY parameters,
$M_2,|\mu|$ and $\tan\beta$, can be extracted 
{\it unambiguously} from the masses $m_{\tilde{\chi}^\pm_{1,2}}$, 
the total production cross sections, and the left-right (LR) asymmetries 
with polarized electron beams, while the phase  $\Phi_\mu$ is
determined up to a twofold ambiguity $\Phi_\mu \leftrightarrow
2 \pi - \Phi_\mu$. [This ambiguity can be resolved by
measuring manifestly CP--noninvariant observables, 
see Ref.~\cite{R11}, related to the normal polarization of the
charginos.]
\s
 
This analysis of the chargino sector is independent of the structure
of the neutralino sector which is potentially more complex than the
form encountered in the 
Minimal Supersymmetric Standard Model (MSSM). The structure of the chargino
sector, by contrast, is isomorphic to the form of the MSSM for a large
class of supersymmetric theories. \s
%

Moreover, from the energy distribution of the final particles
in the decay of the lightest chargino, the mass of the lightest
neutralino can be measured; this allows to determine the other 
U(1) gaugino mass parameter $M_1$ if this parameter 
is real. If not, additional information on the phase of $M_1$
must be derived from observables involving the heavier neutralinos. \s

In summary. If the chargino/neutralino sector is CP--invariant, all 
fundamental gaugino parameters can be derived from the masses
and cross sections of the chargino sector, supplemented by the mass
of the lightest neutralino. In CP--noninvariant theories, the phase
of $\mu$ can be determined up to a twofold ambiguity 
by measuring CP-even observables; the ambiguity can be resolved
by analyzing manifestly CP--noninvariant observables. The phase
of $M_1$ can only be
obtained by exploiting observables involving heavier neutralino states. \s
  
The analysis will be based strictly on low--energy SUSY. Once
these basic parameters will have been extracted experimentally, they may be
confronted, for instance, with the ensemble of  relations predicted in 
Grand Unified Theories. The paper will be divided into six parts. In
Section~2 we recapitulate the central elements of 
the mixing formalism for the charged gauginos and higgsinos. 
In Section~3 the cross sections for chargino production, the left--right 
asymmetries, and the polarization vectors of the charginos are given. 
In Section~4 we describe a phenomenological analysis based on 
a specific mSUGRA scenario to exemplify the procedure for extracting 
the fundamental 
SUSY parameters in a model--independent way. 
In Section~5 we briefly 
comment on the possibility of extracting the U(1) gaugino mass $M_1$ 
from the lightest neutralino mass measured in the decay 
$\tilde{\chi}^\pm_1\rightarrow W^\pm + \tilde{\chi}^0_1$.
Conclusions are given in Section~6.

\subsection*{2. Mixing Formalism}
\label{sec:mixing}
 
Since the chargino mass matrix ${\cal M}_C$ is not symmetric, two
different 
unitary matrices acting on the left-- and right--chiral 
$(\tilde{W},\tilde{H})$ states are needed to diagonalize the matrix:
\begin{eqnarray}
U_{L,R}\left(\begin{array}{c}
             \tilde{W}^- \\
             \tilde{H}^-
             \end{array}\right)_{L,R} =
       \left(\begin{array}{c}
             \tilde{\chi}^-_1 \\
             \tilde{\chi}^-_2
             \end{array}\right)_{L,R} 
\end{eqnarray}
The unitary matrices $U_L$ and $U_R$ can be parametrized in the
following way \cite{R11}:
\begin{eqnarray}
&& U_L=\left(\begin{array}{cc}
             \cos\phi_L & {\rm e}^{-i\beta_L}\sin\phi_L \\
            -{\rm e}^{i\beta_L}\sin\phi_L & \cos\phi_L
             \end{array}\right) \nonumber\\
&& U_R=\left(\begin{array}{cc}
             {\rm e}^{i\gamma_1} & 0 \\
             0 & {\rm e}^{i\gamma_2}
             \end{array}\right)
        \left(\begin{array}{cc}
             \cos\phi_R & {\rm e}^{-i\beta_R}\sin\phi_R \\
            -{\rm e}^{i\beta_R}\sin\phi_R & \cos\phi_R
             \end{array}\right) 
\end{eqnarray}
The eigenvalues $m^2_{\tilde{\chi}^\pm_{1,2}}$ are given by
\begin{eqnarray}
m^2_{\tilde{\chi}^\pm_{1,2}}
   =\frac{1}{2}\left[M^2_2+|\mu|^2+2m^2_W\mp \Delta_C\right]
\end{eqnarray}
with $\Delta_C$ involving the phase  $\Phi_\mu$: 
\begin{eqnarray}
\Delta_C=\sqrt{(M^2_2-|\mu|^2)^2+4m^4_W\cos^2 2\beta
              +4m^2_W(M^2_2+|\mu|^2)+8m^2_WM_2|\mu|
               \sin2\beta\cos\Phi_\mu}
\end{eqnarray}
The quantity $\Delta_C$ determines the difference of the two chargino 
masses: $\Delta_C = m^2_{\tilde{\chi}^\pm_2} - m^2_{\tilde{\chi}^\pm_1}$.
The four phase angles $\{\beta_L,\beta_R,\gamma_1,\gamma_2\}$
are not independent but can be expressed in terms of  the invariant angle
$\Phi_\mu$:
\begin{eqnarray}
&& \tan\beta_L=-\frac{\sin\Phi_\mu}{\cos\Phi_\mu
                                      +\frac{M_2}{|\mu|}\cot\beta}
   \qquad \hskip 0.6cm
   \tan\beta_R=+\frac{\sin\Phi_\mu}{\cos\Phi_\mu
                                      +\frac{M_2}{|\mu|}\tan\beta}
   \nonumber\\
&& \tan\gamma_1=+\frac{\sin\Phi_\mu}{\cos\Phi_\mu
                   +\frac{M_2(m^2_{\tilde{\chi}^\pm_1}-|\mu|^2)}{
                          |\mu| m^2_W\sin 2\beta}} 
   \qquad
   \tan\gamma_2=-\frac{\sin\Phi_\mu}{\cos\Phi_\mu
                   +\frac{M_2m^2_W\sin 2\beta}{
                         |\mu|(m^2_{\tilde{\chi}^\pm_2}-M^2_2)}}
\label{eq:four phases}
\end{eqnarray}
All four phase angles vanish in CP--invariant theories for which 
$\Phi_\mu \rightarrow 0$ or $\pi$.  The rotation angles 
$\phi_L$ and $\phi_R$ satisfy the relations:
\begin{eqnarray}
&&\cos 2\phi_L=-\frac{M_2^2-|\mu|^2-2m^2_W\cos 2\beta}{\Delta_C}
   \nonumber\\ 
&&\sin 2\phi_L=-\frac{2m_W\sqrt{M^2_2+|\mu|^2+(M^2_2-|\mu|^2)\cos 2\beta
                     +2M_2|\mu|\sin2\beta\cos\Phi_\mu}}{\Delta_C}
   \nonumber
\end{eqnarray}
and
\begin{eqnarray}
&&\cos 2\phi_R=-\frac{M_2^2-|\mu|^2+2m^2_W\cos 2\beta}{\Delta_C}
             \nonumber\\ 
&&\sin 2\phi_R=-\frac{2m_W\sqrt{M^2_2+|\mu|^2-(M^2_2-|\mu|^2)\cos 2\beta
                     +2M_2|\mu|\sin2\beta\cos\Phi_\mu}}{\Delta_C}
\end{eqnarray}
As a consequence of possible field redefinitions, the parameters
$\tan\beta$ and $M_2$ can be chosen  
real and positive. \s

The fundamental SUSY parameters $M_2$, $|\mu|$, $\tan\beta$ 
and the phase parameter $\cos\Phi_\mu$ can be extracted from the
chargino
$\tilde{\chi}^\pm_{1,2}$ parameters: the masses 
$m_{\tilde{\chi}^\pm_{1,2}}$ and the two mixing angles 
$\phi_L$ and $\phi_R$ of the left-- and
right--chiral components of the wave function. These mixing angles are
physical observables and they can be measured, as well as the chargino
masses
$m_{\tilde{\chi}^\pm_{1,2}}$,  in the processes
$e^+e^-\rightarrow\tilde{\chi}^-_i\tilde{\chi}^+_j \, \, \, \, [i,j=1,2]$.

The two angles $\phi_L$ and $\phi_R$ and the nontrivial phase angles
$\{\beta_L,\beta_R,\gamma_1,\gamma_2\}$ define the couplings of the 
chargino--chargino--$Z$ vertices and the electron--sneutrino--chargino
vertex:
\begin{eqnarray}
\langle\tilde{\chi}^-_{1L}|Z|\tilde{\chi}^-_{1L}\rangle 
 &=& -\frac{e}{s_W c_W} \left[s_W^2 - \frac{3}{4}-\frac{1}{4}
     \cos 2\phi_L\right] \nonumber\\
\langle\tilde{\chi}^-_{1L}|Z|\tilde{\chi}^-_{2L}\rangle 
 &=& +\frac{e}{4s_W c_W}{\rm e}^{-i\beta_L}\sin 2\phi_L \nonumber\\
\langle\tilde{\chi}^-_{2L}|Z|\tilde{\chi}^-_{2L}\rangle 
 &=& -\frac{e}{s_W c_W} \left[s_W^2 - \frac{3}{4}+\frac{1}{4}
     \cos 2\phi_L\right] \nonumber\\
\langle\tilde{\chi}^-_{1R}|Z|\tilde{\chi}^-_{1R}\rangle 
 &=& -\frac{e}{s_W c_W} \left[s_W^2-\frac{3}{4}-\frac{1}{4}\cos 
    2\phi_R\right] \nonumber\\
\langle\tilde{\chi}^-_{1R}|Z|\tilde{\chi}^-_{2R}\rangle 
 &=& +\frac{e}{4s_W c_W}{\rm e}^{-i(\beta_R-\gamma_1+\gamma_2)}
                        \sin 2\phi_R\nonumber\\
\langle\tilde{\chi}^-_{2R}|Z|\tilde{\chi}^-_{2R}\rangle 
 &=& -\frac{e}{s_W c_W} \left[s_W^2-\frac{3}{4}+\frac{1}{4}\cos 
    2\phi_R\right] \nonumber\\
\langle\tilde{\chi}^-_{1R}|\tilde{\nu}|e^-_L\rangle 
 &=& -\frac{e}{s_W}{\rm e}^{i\gamma_1}\cos\phi_R \nonumber\\
\langle\tilde{\chi}^-_{2R}|\tilde{\nu}|e^-_L\rangle 
 &=& +\frac{e}{s_W}{\rm e}^{i(\beta_R+\gamma_2)}\sin\phi_R
\label{eq:vertex}
\end{eqnarray}
where $s_W^2 =1-c_W^2 \equiv \sin^2\theta_W$. The coupling to the
higgsino component, being proportional to the electron mass, has been
neglected in the sneutrino vertex; the sneutrino couples only to
left--handed electrons. 
Note that the CP--noninvariant phase $\Phi_\mu$ 
enters the vertices through the phase angles which have been expressed 
in terms of the fundamental SUSY parameters in eq.(\ref{eq:four phases}).
Since the photon--chargino vertex is diagonal,
it does not depend on the mixing angles:
\begin{eqnarray}
\langle\tilde{\chi}^-_{iL,R}|\gamma|\tilde{\chi}^-_{jL,R}\rangle 
   = e \delta_{ij} 
\end{eqnarray}
The parameter $e$ is the electromagnetic coupling which will   
be taken at an effective scale identified with the c.m. energy $\sqrt{s}$.

\subsection*{3. Chargino Pair--Production}
\label{sec:production}

The process $e^+e^-\rightarrow\tilde{\chi}^-_i\tilde{\chi}^+_j$ is
generated
by the three mechanisms shown in Fig.1: $s$--channel $\gamma$ and $Z$
exchanges, and $t$--channel $\tilde{\nu}$ exchange. The transition matrix
element, after a Fierz transformation of the $\tilde{\nu}$--exchange
amplitude,
\begin{eqnarray}
T\left(e^+e^-\rightarrow\tilde{\chi}^-_i\tilde{\chi}^+_j\right)
 = \frac{e^2}{s}Q_{\alpha\beta}
   \left[\bar{v}(e^+)  \gamma_\mu P_\alpha  u(e^-)\right]
   \left[\bar{u}(\tilde{\chi}^-_i) \gamma^\mu P_\beta 
               v(\tilde{\chi}^+_j) \right]
\label{eq:production amplitude}
\end{eqnarray}
can be expressed in terms of four bilinear charges, defined by  
the chiralities $\alpha,\beta=L,R$ of the associated 
lepton and chargino currents\\ \s 
\noindent
{\bf (i)} \underline{$\tilde{\chi}^-_1\tilde{\chi}^+_1$}
\begin{eqnarray}
Q_{LL}&=&1+ \frac{D_Z}{s_W^2 c_W^2}(s_W^2 -\frac{1}{2}) 
         \left(s_W^2 -\frac{3}{4}-\frac{1}{4}\cos 2\phi_L\right) 
         \nonumber\\ 
Q_{LR}&=&1+ \frac{D_Z}{s_W^2 c_W^2} (s_W^2 -\frac{1}{2}) 
         \left(s_W^2-\frac{3}{4}-\frac{1}{4}\cos 2\phi_R\right) 
        + \frac{D_{\tilde{\nu}}}{4s_W^2} (1+\cos 2\phi_R) \nonumber\\
Q_{RL}&=&1+\frac{D_Z}{c_W^2} \left(s_W^2 -\frac{3}{4}-\frac{1}{4}\cos 
          2\phi_L\right) \nonumber\\
Q_{RR}&=&1+ \frac{D_Z}{c_W^2}  \left(s_W^2 -\frac{3}{4}-\frac{1}{4}\cos 
         2\phi_R\right)
\end{eqnarray}\\
\noindent
{\bf (ii)} \underline{$\tilde{\chi}^-_1\tilde{\chi}^+_2$}
\begin{eqnarray}
Q_{LL}&=&\frac{D_Z}{4 s_W^2 c_W^2}(s_W^2 -\frac{1}{2})
         {\rm e}^{-i\beta_L}\sin 2\phi_L 
         \nonumber\\ 
Q_{LR}&=&\frac{D_Z}{4 s_W^2 c_W^2} (s_W^2 -\frac{1}{2})
          {\rm e}^{-i(\beta_R-\gamma_1+\gamma_2)}\sin 2\phi_R 
         +\frac{D_{\tilde{\nu}}}{4s_W^2}
          {\rm e}^{-i(\beta_R-\gamma_1+\gamma_2)}\sin 2\phi_R \nonumber\\
Q_{RL}&=&\frac{D_Z}{4 c_W^2}{\rm e}^{-i\beta_L}\sin 2\phi_L \nonumber\\
Q_{RR}&=&\frac{D_Z}{4 c_W^2}{\rm e}^{-i(\beta_R-\gamma_1+\gamma_2)}
                             \sin 2\phi_R
\end{eqnarray}
\noindent
{\bf (iii)} \underline{$\tilde{\chi}^-_2\tilde{\chi}^+_2$}
\begin{eqnarray}
Q_{LL}&=&1+ \frac{D_Z}{s_W^2 c_W^2}(s_W^2 -\frac{1}{2}) 
         \left(s_W^2 -\frac{3}{4}+\frac{1}{4}\cos 2\phi_L\right) 
         \nonumber\\ 
Q_{LR}&=&1+ \frac{D_Z}{s_W^2 c_W^2} (s_W^2 -\frac{1}{2}) 
         \left(s_W^2-\frac{3}{4}+\frac{1}{4}\cos 2\phi_R\right) 
        + \frac{D_{\tilde{\nu}}}{4s_W^2} (1-\cos 2\phi_R) \nonumber\\
Q_{RL}&=&1+\frac{D_Z}{c_W^2} \left(s_W^2 -\frac{3}{4}+\frac{1}{4}\cos 
          2\phi_L\right) \nonumber\\
Q_{RR}&=&1+ \frac{D_Z}{c_W^2}  \left(s_W^2 -\frac{3}{4}+\frac{1}{4}\cos 
         2\phi_R\right)
\end{eqnarray}\\
The first index in $Q_{\alpha \beta}$ refers to the chirality of the
$e^\pm$ 
current, the second index to the chirality of the $\tilde{\chi}_1^\pm$ 
current. The $\tilde{\nu}$ 
exchange affects only the $LR$ chirality charge while all other amplitudes 
are built up by $\gamma$ and $Z$ exchanges only. $D_{\tilde{\nu}}$ denotes
the sneutrino propagator
$D_{\tilde{\nu}} = s/(t- m_{\tilde{\nu}}^2)$, and 
$D_Z$ the $Z$ propagator $D_Z=s/(s-m^2_Z+im_Z\Gamma_Z)$; the non--zero 
$Z$ width can in general be neglected for the energies considered 
in the present analysis so that the charges are rendered complex in 
the present Born
approximation only through the CP--noninvariant phases.  \s 

For the sake of convenience we also introduce the eight quartic charges
\cite{R7}
defined in Table~\ref{quartic:charges}. These charges
correspond to the eight independent helicity amplitudes describing 
the chargino
production processes for massless electrons/positrons. \s
\begin{table}[\hbt]
\renewcommand{\arraystretch}{1.5}
\begin{center}
\begin{tabular} {|c|c|l|}\hline \hline
${\cal P}$ & ${\cal CP}$ & \hskip 2cm Quartic Charges \\ \hline \hline
 {      }  &  {       }  & $Q_1=\frac{1}{4}\left[|Q_{RR}|^2+|Q_{LL}|^2
                               +|Q_{RL}|^2+|Q_{LR}|^2\right]$ \\
   even    &    even     & $Q_2=\frac{1}{2}{\rm Re}\left[Q_{RR}Q^*_{RL}
                               +Q_{LL}Q^*_{LR}\right]$ \\
 {      }  &  {       }  & $Q_3=\frac{1}{4}\left[|Q_{RR}|^2+|Q_{LL}|^2
                          -|Q_{RL}|^2-|Q_{LR}|^2\right]$ \\ 
                           \cline{2-3}
           &    odd      & $Q_4=\frac{1}{2}  {\rm Im}\left[Q_{RR}Q^*_{RL}
                               +Q_{LL}Q^*_{LR}\right]$ \\ \hline \hline
 {      }  &  {       }  & $Q'_1=\frac{1}{4}\left[|Q_{RR}|^2+|Q_{RL}|^2
                               -|Q_{LR}|^2-|Q_{LL}|^2\right]$ \\
   odd     &    even     & $Q'_2=\frac{1}{2}{\rm Re}\left[Q_{RR}Q^*_{RL}
                               -Q_{LL}Q^*_{LR}\right]$ \\
 {      }  &  {       }  & $Q'_3=\frac{1}{4}\left[|Q_{RR}|^2+|Q_{LR}|^2
                               -|Q_{RL}|^2-|Q_{LL}|^2\right]$ \\ 
                           \cline{2-3}
           &    odd      & $Q'_4=\frac{1}{2}{\rm Im}\left[Q_{RR}Q^*_{RL}
                               -Q_{LL}Q^*_{LR}\right]$ \\ \hline \hline
\end{tabular}
\renewcommand{\arraystretch}{1.2}
\caption{\label{quartic:charges}
{\it Quartic charges determining the cross section and polarization
vectors in pair production of charginos in $e^+e^-$ collisions. Detailed
comments are given in the text.}}
\end{center}
\end{table}

The charges $Q_1$ to $Q_4$ are manifestly parity--even, i.e. invariant
under space reflection; $Q'_1$ to $Q'_4$ are parity--odd. The charges 
$Q_1$ to $Q_3$ and $Q'_1$ to $Q'_3$ are CP
invariant\footnote[1]
{When expressed in terms of the fundamental SUSY parameters, they do
depend nevertheless indirectly on $\cos\Phi_\mu$ 
through $\cos 2\phi_{L,R}$, in
the same way as the masses depend indirectly on this parameter.} while
$Q_4$
and $Q'_4$ change sign under CP transformations. 
The CP invariance of $Q_2$ and $Q'_2$ can
easily be shown by noting that
\begin{eqnarray}
\cos(\beta_L-\beta_R+\gamma_1-\gamma_2)\sin 2\phi_L \sin 2\phi_R
\makebox[5cm]{} \nonumber  \\             
\makebox[3cm]{}    =
\frac{m^2_{\tilde{\chi}^\pm_1}+m^2_{\tilde{\chi}^\pm_2}}{
       2m_{\tilde{\chi}^\pm_1} m_{\tilde{\chi}^\pm_2}}
\left(1-\cos 2\phi_L \cos 2\phi_R\right)
-\frac{2m^2_W}{m_{\tilde{\chi}^\pm_1} m_{\tilde{\chi}^\pm_2}} 
\label{eq:sisisi} 
\end{eqnarray}
Thus all the cross sections 
$e^+e^-\rightarrow \tilde{\chi}^-_i\tilde{\chi}^+_j$ for any combination
of pairs $(ij)$ depend
only on $\cos 2 \phi_L$ and $\cos 2 \phi_R$ altogether. For polarized 
electron beams the sums and differences of the quartic charges are
restricted 
to either $L$ or $R$ components (first index) of the $e^\pm$ currents.\s 

The measurement of the quartic charges $Q_1$ to $Q'_3$ in the
total cross sections and left--right asymmetries for equal
and mixed chargino pair--production allows us to extract 
 the two terms $\cos 2\phi_L$ and $\cos 2\phi_R$ unambiguously
as will be demonstrated explicitly in the following section. \s

The CP--noninvariant charges $Q_4$ and $Q'_4$ vanish for equal 
chargino pairs $\tilde{\chi}^-_1\tilde{\chi}^+_1$ and 
$\tilde{\chi}^-_2\tilde{\chi}^+_2$. They can be determined only by
measuring observables related to 
the normal components of the $\tilde{\chi}^\pm_{1,2}$ polarization vectors 
in mixed $e^+e^- \rightarrow\tilde{\chi}^-_1\tilde{\chi}^+_2 / 
\tilde{\chi}^-_2 \tilde{\chi}^+_1$ pair production \cite{R11}.\s
\vspace*{1cm} 
\begin{center}
\begin{picture}(330,100)(0,0)
\Text(15,85)[]{$e^-$}
\ArrowLine(10,75)(35,50)
\ArrowLine(35,50)(10,25)
\Text(15,15)[]{$e^+$}
\Photon(35,50)(75,50){4}{8}
\Text(55,37)[]{$\gamma$}
\ArrowLine(75,50)(100,75)
\Photon(75,50)(100,75){3}{7}
\Text(95,85)[]{$\tilde{\chi}^-_i$}
\ArrowLine(100,25)(75,50)
\Photon(100,25)(75,50){3}{7}
\Text(95,15)[]{$\tilde{\chi}^+_j$}
\Text(125,85)[]{$e^-$}
\ArrowLine(120,75)(145,50)
\Text(125,15)[]{$e^+$}
\ArrowLine(145,50)(120,25)
\Photon(145,50)(185,50){4}{8}
\Text(165,37)[]{$Z$}
\ArrowLine(185,50)(210,75)
\Photon(185,50)(210,75){3}{7}
\Text(207,85)[]{$\tilde{\chi}^-_i$}
\ArrowLine(210,25)(185,50)
\Photon(210,25)(185,50){3}{7}
\Text(207,15)[]{$\tilde{\chi}^+_j$}
\Text(235,85)[]{$e^-$}
\ArrowLine(230,75)(275,75)
\Text(235,15)[]{$e^+$}
\ArrowLine(275,25)(230,25)
\Line(274,75)(274,25)
\Line(276,75)(276,25)
\Text(285,50)[]{$\tilde{\nu}$}
\ArrowLine(275,75)(320,75)
\Photon(275,75)(320,75){3}{7}
\Text(318,85)[]{$\tilde{\chi}^-_i$}
\ArrowLine(320,25)(275,25)
\Photon(320,25)(275,25){3}{7}
\Text(318,15)[]{$\tilde{\chi}^+_j$}
\end{picture}\\
\end{center}
\smallskip

Figure~1: {\it The three exchange mechanisms contributing to the 
            production of chargino  \\ \hspace*{0.41cm} 
            pairs $\tilde{\chi}^-_i \tilde{\chi}^+_j$ in $\ee$
            annihilation.}
\bigskip 

Defining the $\tilde{\chi}^-_i$ production angle with respect to the
electron flight--direction by $\Theta$, the helicity amplitudes can be
derived from eq.(\ref{eq:production amplitude}). While electron
and positron helicities are opposite to each other in all amplitudes,
the $\tilde{\chi}^-_i$ and $\tilde{\chi}^+_j$ helicities are in
general not correlated due to the non--zero masses of the particles;
amplitudes with equal $\tilde{\chi}^-_i$ and $\tilde{\chi}^+_j$
helicities vanish only $\propto m_{\tilde{\chi}^\mp_{i,j}} /\sqrt{s}$ 
for asymptotic energies.
Denoting the electron helicity by the first index, the
$\tilde{\chi}^-_i$ and $\tilde{\chi}^+_j$ helicities by the remaining
two indices, $\lambda_i$ and $\lambda_j$, respectively, 
the helicity amplitudes $T(\sigma;\lambda_i,
\lambda_j)=2\pi\alpha\langle\sigma;\lambda_i\lambda_j\rangle$
are given as follows \cite{R8},
\begin{eqnarray}
&& \langle +;++\rangle 
   =-\left[Q_{RR}\sqrt{1-\eta^2_+}+Q_{RL}\sqrt{1-\eta^2_-}\right]
           \sin\Theta \nonumber\\
&& \langle +;+-\rangle 
   =-\left[Q_{RR}\sqrt{(1+\eta_+)(1+\eta_-)}
          +Q_{RL}\sqrt{(1-\eta_+)(1-\eta_-)}\right]
          (1+\cos\Theta) \nonumber\\
&& \langle +;-+\rangle 
   =+\left[Q_{RR}\sqrt{(1-\eta_+)(1-\eta_-)}
          +Q_{RL}\sqrt{(1+\eta_+)(1+\eta_-)}\right]
          (1-\cos\Theta) \nonumber\\
&& \langle +;--\rangle 
   =+\left[Q_{RR}\sqrt{1-\eta^2_-}+Q_{RL}\sqrt{1-\eta^2_+}\right]
           \sin\Theta 
\end{eqnarray}
and
\begin{eqnarray}
&& \langle -;++\rangle 
   =-\left[Q_{LR}\sqrt{1-\eta^2_+}+Q_{LL}\sqrt{1-\eta^2_-}\right]
           \sin\Theta \nonumber\\
&& \langle -;+-\rangle 
   =+\left[Q_{LR}\sqrt{(1+\eta_+)(1+\eta_-)}
          +Q_{LL}\sqrt{(1-\eta_+)(1-\eta_-)}\right]
          (1-\cos\Theta) \nonumber\\
&& \langle -;-+\rangle 
   =-\left[Q_{LR}\sqrt{(1-\eta_+)(1-\eta_-)}
          +Q_{LL}\sqrt{(1+\eta_+)(1+\eta_-)}\right]
          (1+\cos\Theta) \nonumber\\
&& \langle -;--\rangle 
   =+\left[Q_{LR}\sqrt{1-\eta^2_-}+Q_{LL}\sqrt{1-\eta^2_+}\right]
           \sin\Theta 
\label{eq:helicity amplitude}
\end{eqnarray}
where $\eta_\pm=\lambda^{1/2}(1,\mu^2_i, \mu^2_j)\pm(\mu^2_i-\mu^2_j)$ with
the 2-body phase--space function 
$\lambda(1,\mu^2_i, \mu^2_j)=[1-(\mu_i+\mu_j)^2]
[1-(\mu_i-\mu_j)^2]$ and the reduced masses
$\mu^2_i=m^2_{\tilde{\chi}^\pm_i}/s$. From these amplitudes the 
$\tilde{\chi}^-_i \tilde{\chi}^+_j$ production cross sections and the 
left--right  asymmetries can be determined. \s

\subsubsection*{3.1 Production cross sections}

The unpolarized differential cross section is given by the average/sum
over the helicities:
\begin{eqnarray}
\frac{{\rm d}\sigma}{{\rm d}\cos\Theta}
      (e^+e^-\rightarrow\tilde{\chi}^-_i\tilde{\chi}^+_j)
 =\frac{\pi\alpha^2}{32 s} \lambda^{1/2} \, 
  \sum_{\sigma\lambda_i\lambda_j}\,
  |\langle\sigma;\lambda_i\lambda_j\rangle|^2
\end{eqnarray}
where $\lambda$ is the two--body phase space function introduced
above.
Carrying out the sum,  the following expression for the cross 
section in terms of the quartic charges can be derived:
\begin{eqnarray}
&& \frac{{\rm d}\sigma}{{\rm d}\cos\Theta}
      (e^+e^-\rightarrow\tilde{\chi}^-_i\tilde{\chi}^+_j)\nonumber\\
&&{ } \hskip 0.5cm  =\frac{\pi\alpha^2}{2 s}\lambda^{1/2} 
  \bigg\{\left[1-(\mu^2_i - \mu^2_j)^2+\lambda\cos^2\Theta\right]Q_1
          +4\mu_i\mu_j Q_2+2\lambda^{1/2} Q_3 \cos\Theta \bigg\}
\label{eq:cross section}
\end{eqnarray}
If the production angle could be measured unambiguously 
on an event--by--event basis, the quartic charges could be extracted
directly from the angular dependence of the cross section at a single 
energy. After integration over the production angle $\Theta$, the total
cross section still depends on $Q_3$ since $t$--channel 
sneutrino exchange gives rise to a non--linear 
forward--backward asymmetric angular dependence.\s  

The total production cross section is shown in Fig.~2 as a function of  
the c.m.~energy for a fixed sneutrino mass. The sneutrino mass is assumed
to be predetermined from direct production $e^+e^-\rightarrow\tilde{\nu}_e 
\overline{\tilde{\nu}_e}$. The curves, which should be interpreted as
characteristic examples, are based on the two 
CP--invariant 
mSUGRA scenarios introduced in Ref.~\cite{LCWS}. They correspond to a 
small and a large $\tan\beta$ solution for the universal gaugino and 
scalar masses:
\begin{eqnarray}
\begin{array}{lll}
\mbox{\boldmath $RR1$}: & {\rm small}\ \ \tan\beta\ \ 
=\hskip 0.2cm 3 \makebox[1mm]{}:
 & (m_0,M_{\frac{1}{2}})=(100\  {\rm GeV}, 200\  {\rm GeV})\ \
   \\
\mbox{\boldmath $RR2$}: & {\rm large}\ \ \hskip 0.1cm\tan\beta\ \ =30 
\makebox[1mm]{}: 
 & (m_0,M_{\frac{1}{2}})=(160\  {\rm GeV}, 200\  {\rm GeV})\ \
\label{eq:parameter}
\end{array}
\end{eqnarray}
The induced chargino $\tilde{\chi}^0_{1,2}$, neutralino 
$\tilde{\chi}^0_1$ and sneutrino masses $\tilde{\nu}$ are 
collected in Table~\ref{mSUGRA}. The CP-phase
$\Phi_\mu$ is set to zero.
\begin{table}[\hbt]
\renewcommand{\arraystretch}{1.5}
\begin{center}
\begin{tabular} {c|c|cl|}\hline \hline
$\tilde{m}$ [GeV]    & \mbox{\boldmath $RR1$} : $\tan\beta=3$ & 
\mbox{\boldmath $RR2$} : $\tan\beta=30$ \\
\hline \hline
$M_2$                & 152                 & 150   \\
$\mu$                & 316                 & 263   \\ \hline
$\tilde{\chi}^\pm_1$ & 128                 & 132   \\
$\tilde{\chi}^\pm_2$ & 346                 & 295   \\ \hline
$\tilde{\chi}^0_1$   & 70                  & 72    \\ \hline
$\tilde{\nu}$        & 166                 & 206   \\ \hline \hline
\end{tabular}
\renewcommand{\arraystretch}{1.2}
\caption{\it Gaugino and higgsino mass parameters, mass values of 
             the charginos and the lightest neutralino, and of
             the sneutrino in the reference points of the mSUGRA 
             scenarios introduced in Ref.~\cite{LCWS}.}
\label{mSUGRA}
\end{center}
\end{table}
The sharp rise of the production cross sections in Fig.~2  
allows to measure  the chargino mass $m_{\tilde{\chi}^\pm_{1,2}}$ very 
precisely \cite{R2A,R5}. 
Fig.~3 exhibits the angular distribution as a function of the 
scattering angle for the  parameters of Table~\ref{mSUGRA}
at the c.m. energy  800 GeV.   The peak in the near--forward 
region is due to the $t$-channel sneutrino exchange.

\subsubsection*{3.2 Left-right asymmetries}
\label{subsec:left-right asymmetry}

Switching the longitudinal electron polarization yields
a left-right (LR) asymmetry ${\cal A}_{LR}$, defined as
\begin{eqnarray}
{\cal A}_{LR} = \frac{1}{4}\sum_{\lambda_i\lambda_j}
       \bigg[|\langle +;\lambda_i\lambda_j\rangle|^2
            -|\langle -;\lambda_i\lambda_j\rangle|^2\bigg]/{\cal N}
\end{eqnarray}
with the normalization
\begin{eqnarray}
{\cal N}=\frac{1}{4}\sum_{\lambda_i\lambda_j}
         \bigg[|\langle +;\lambda_i\lambda_j\rangle|^2
              +|\langle -;\lambda_i\lambda_j\rangle|^2\bigg]
\end{eqnarray}
The LR asymmetry ${\cal A}_{LR}$ can be readily expressed in terms of the
quartic charges,
\begin{eqnarray}
{\cal A}_{LR}=4 \bigg\{[1-(\mu^2_i-\mu^2_j)^2+\lambda\cos^2\Theta]Q'_1
       +4\mu_i\mu_j Q'_2+2\lambda^{1/2}\cos\Theta Q'_3\bigg\} / {\cal N}
\end{eqnarray}
with, correspondingly,
\begin{eqnarray}
{\cal N}=4\bigg\{[1-(\mu^2_i-\mu^2_j)^2+\lambda\cos^2\Theta]Q_1
       +4\mu_i\mu_j Q_2+2\lambda^{1/2}\cos\Theta Q_3\bigg\}
\end{eqnarray}

In Fig.~4  the LR asymmetries are depicted as a function of the 
scattering angle for the parameters of Table~\ref{mSUGRA}
at the c.m. energy  800 GeV. The large negative asymmetry for 
$\tilde{\chi}^-_1 \tilde{\chi}^+_1$ production in forward--direction is
due to the $t$-channel sneutrino exchange which affects only the cross 
section for left-handed electron beams.

\subsubsection*{3.3 Polarization vectors}

The polarization vector $\vec{\cal P}=({\cal P}_L,{\cal P}_T,
{\cal P}_N)$ is defined in the rest frame\footnote[2]{Axis $\hat{z}\| L$ 
in the flight direction of $\tilde{\chi}^-_i$, $\hat{x}\| T$ rotated 
counter--clockwise in the production plane, and 
$\hat{y}=\hat{z}\times\hat{x}\| N$.}
of the chargino 
$\tilde{\chi}^-_i$. ${\cal P}_L$ denotes the component parallel to the
$\tilde{\chi}^-_i$ flight direction in the c.m. frame,
${\cal P}_T$ the transverse component in the production plane, and 
${\cal P}_N$ the component normal to the production plane. 
These three components can be expressed by helicity amplitudes in
the following way:
\begin{eqnarray}
&& {\cal P}_L=\frac{1}{4}\sum_{\sigma=\pm}\left\{
              |\langle\sigma;++\rangle|^2+|\langle\sigma;+-\rangle|^2
             -|\langle\sigma;-+\rangle|^2-|\langle\sigma;--\rangle|^2
                                           \right\}/{\cal N}
              \nonumber\\
&& {\cal P}_T=\frac{1}{2}{\rm Re}\bigg\{\sum_{\sigma=\pm}\left[
              |\langle\sigma;++\rangle\langle\sigma;-+\rangle^*
             +|\langle\sigma;--\rangle\langle\sigma;+-\rangle^*
                          \right]\bigg\}/{\cal N}\nonumber\\
&& {\cal P}_N=\frac{1}{2}{\rm Im}\bigg\{\sum_{\sigma=\pm}\left[
              |\langle\sigma;--\rangle\langle\sigma;+-\rangle^*
             -|\langle\sigma;++\rangle\langle\sigma;-+\rangle^*
                           \right]\bigg\}/{\cal N}
\end{eqnarray}
%
%
%
%
%
The longitudinal, transverse and normal components of the 
$\tilde{\chi}^-_i$ polarization vector can easily be obtained from the 
helicity amplitudes. Expressed in terms of the quartic charges, they
read:
\begin{eqnarray}
&& {\cal P}_L = 4 \bigg\{2(1-\mu^2_i-\mu^2_j)\cos\Theta Q'_1
       +4\mu_i\mu_j\cos\Theta Q'_2\nonumber\\
&& { } \hskip 3cm +\lambda^{1/2}[1+\cos^2\Theta
         -(\mu^2_i-\mu^2_j)\sin^2\Theta]Q'_3\bigg\}/{\cal N}
         \nonumber\\                          
&& {\cal P}_T = -8\bigg\{ [(1-\mu^2_i+\mu^2_j) Q'_1
       +\lambda^{1/2}Q'_3\cos\Theta]\mu_i
       +(1+\mu^2_i-\mu^2_j)\mu_j Q'_2\bigg\}\sin\Theta/{\cal N}
         \nonumber\\
&& {\cal P}_N = 8\lambda^{1/2}\mu_j\sin\Theta Q_4/{\cal N}
\label{eq:polarization vector}
\end{eqnarray}
The longitudinal and transverse components are P--odd and CP--even, and 
the normal component is P--even and CP--odd.\s 

The normal polarization component can only be generated by complex 
production amplitudes, c.f. Ref.\cite{Norm}.
Non--zero phases are present in the fundamental 
SUSY parameters if CP is broken in the supersymmetric 
interaction. Also the non--zero width of the $Z$ boson and loop 
corrections generate non--trivial phases; 
however, the width effect is negligible for high energies and the effects
due to radiative corrections are small as well. So, the normal component
is effectively generated by the complex SUSY couplings. The bilinear charges
are real in the diagonal modes (1,1) and (2,2) so that 
the normal polarization vanishes.
But, the non--diagonal modes (1,2) and (2,1) 
may have non-vanishing normal 
polarization components, determined by the quartic charge
\begin{eqnarray}
Q_4&=&\frac{1}{32c^4_W s^4_W}
      \left[D^2_Z(2s^4_W-s^2_W+\frac{1}{4})
           +D_Z D_{\tilde{\nu}} c^2_W(s^2_W-\frac{1}{2})\right]\nonumber\\
    &&{ }\hskip 1.5cm \times\sin 2\phi_L\sin 2\phi_R
                    \sin(\beta_L-\beta_R+\gamma_1-\gamma_2)
\end{eqnarray}           
When combined with the relation in eq.(\ref{eq:sisisi}), the unknown 
sign of the product $\sin 2\phi_L$ with $\sin 2\phi_R$ can be eliminated. 
The ensuing coefficient $\tan(\beta_L-\beta_R+\gamma_1-\gamma_2)$ 
depends on  $\sin\Phi_\mu$ as evident from  the definition
of the four phase angles $\{\gamma_1,\gamma_2,\beta_L,\beta_R\}$ in
eq.(\ref{eq:four phases}). The normal polarization component is 
generally small. Since CP violating effects like ${\cal P}_N$ or $Q_4$
are proportional to the imaginary part of $M_2\mu m^2_W\sin 2\beta$, 
i.e. the product of 
the ${\cal M}_C$ matrix elements, they vanish for asymptotically large
values of $\tan\beta$. A few numerical examples are
displayed for $\sqrt{s}=800$ GeV in Table.~\ref{tab:normal}, based on
the two reference points \mbox{\boldmath $RR1$} and \mbox{\boldmath $RR2$}
introduced earlier, and the CP phase $\Phi_\mu=\pi/2$.

\begin{table}[\hbt]
\renewcommand{\arraystretch}{1.5}
\begin{center}
\begin{tabular} {|c|c|c|c|}\hline \hline
 {    }        &  $\Theta$         &  $Q_4$   &  ${\cal P}_N$ \\ \hline \hline
 \mbox{\boldmath $RR1$}  & $\pi/4 $  & $-0.199$ &   $-0.333$    \\ \cline{2-4}
 {    }        & $\pi/2 $  & $-0.073$ &   $-0.246$    \\ \cline{2-4}
 {    }        & $3\pi/4$  & $-0.044$ &   $-0.129$    \\ \hline \hline
 \mbox{\boldmath $RR2$}  & $\pi/4 $  & $-0.026$ &   $-0.043$    \\ \cline{2-4}
 {    }        & $\pi/2 $  & $-0.010$ &   $-0.027$    \\ \cline{2-4}
 {    }        & $3\pi/4$  & $-0.006$ &   $-0.013$    \\ \hline \hline
\end{tabular}
\renewcommand{\arraystretch}{1.2}
\caption{\label{tab:normal}
{\it Values of the CP--odd quartic charge $Q_4$ and the normal polarization
     component ${\cal P}_N$ for $\sqrt{s}=800$ GeV and three production 
     angles $\Theta$. The reference points \mbox{\boldmath $RR1/2$} have
     been defined earlier; the CP angle $\Phi_\mu$ is chosen $\pi/2$. }}
\end{center}
\end{table}
%

Due to the two escaping LSP's, it is difficult to measure the 
normal polarization components. 
Nevertheless, CP--odd observables, that are indirectly related to $Q_4$ and
${\cal P}_N$, may be constructed to measure the sign of \
$\sin\Phi_\mu$ -- the only parameter left to be determined. 
An example is the triple product of the initial
electron momentum and the two final--state lepton momenta in the 
$\tilde{\chi}^\pm_{1,2}$ leptonic decays, 
${\cal O}_3={\rm sgn}\left[\vec{p}_{e^-}\cdot 
(\vec{p}_{l^-} \times \vec{p}_{l^+})\right]$. This observable depends on the 
phenomenological analysis powers $\kappa_1$ and $\bar{\kappa}_2$ which 
however can be measured experimentally; therefore, the analysis does not 
require any knowledge of the structure of the neutralino sector.
In particular, the observable ${\cal O}_3$,
based on single--particle momenta
of the two parent charginos, does not depend on potentially
CP-violating couplings in the decay processes, c.f. Ref.\cite{Bern}. \s

\subsection*{4. Observables and Extraction of SUSY Parameters}
\label{sec:observable}
\subsubsection*{4.1 Phenomenological analysis}
 
The pair production of the charginos $\chi^-_i$ and $\tilde{\chi}^+_j$
is characterized by the chargino masses $m_{\tilde{\chi}^\pm_{1,2}}$
and the two mixing 
angles, $\phi_L$ and $\phi_R$ [besides the sneutrino mass
$m_{\tilde{\nu}}$].
These  quantities can be determined 
from three chargino pair--production cross sections and three 
LR asymmetries. Nevertheless, we assume the sneutrino mass 
to be  measured
independently in  sneutrino pair--production.\s
 
The chargino masses $m_{\tilde{\chi}^\pm_{1,2}}$ can be 
determined very precisely at the per--mille level near the threshold
where the production cross sections 
$\sigma(e^+e^-\rightarrow \tilde{\chi}^-_1\tilde{\chi}^+_1)$,
$\sigma(e^+e^-\rightarrow \tilde{\chi}^-_1\tilde{\chi}^+_2)$
and
$\sigma(e^+e^-\rightarrow \tilde{\chi}^-_2\tilde{\chi}^+_2)$
rise sharply with the chargino velocities.\s 

Combining the energy variation of the cross sections with the measurements 
of the LR asymmetries, the two mixing angles $\phi_L$, $\phi_R$ and 
$\cos\Phi_\mu$ can be extracted. 
Based on the first parameter set in eq.(\ref{eq:parameter}),
we will demonstrate that the three chargino production modes enable 
us to extract {\it unambiguously} the values of two cosines, 
$\cos 2\phi_L$ and $\cos 2\phi_R$, by measuring only their production 
cross sections and LR asymmetries with longitudinally-polarized 
electron beams. 
In the mSUGRA scenario implemented with radiative corrections,
the parameter set (\ref{eq:parameter}) with $\tan\beta =3$ 
leads to the following values for cross 
sections and asymmetries at the c.m. energy $\sqrt{s} = 800$ GeV:
\begin{eqnarray}
& \mbox{\boldmath $RR1$}: & \sigma_{tot}(1,1)  \ = 0.197{\rm pb},\ \
   \sigma_{tot}(1,2) = 0.068{\rm pb},\ \
   \sigma_{tot}(2,2) = 0.101{\rm pb}\nonumber\\
&& A_{LR}(1,1) = -0.995,\ \
   A_{LR}(1,2) = -0.911,\ \
   A_{LR}(2,2) = -0.668
\label{eq:measured}
\end{eqnarray}
From now on we will interpret this set as 
experimentally ``measured values", neglecting  experimental errors
for the time being.
The set\footnote[3]{The large $\tan\beta$ set in (\ref{eq:parameter}) 
leads to the same conclusions.}
will be exploited to pin down a unique point in the 
$\{\cos 2\phi_L,\cos 2\phi_R\}$ plane which leads back, 
in combination with the masses, to a unique solution for the fundamental
SUSY parameters.\s

Fig.~5 exhibits the contours in the $\{\cos 2\phi_L,\cos 2\phi_R\}$
plane for the ``measured values" of the
cross sections, $\sigma_{tot}(1,1)$,  $\sigma_{tot}(1,2)$ and
$\sigma_{tot}(2,2)$ 
and the LR asymmetries, $A_{LR}(1,1)$, $A_{LR}(1,2)$ and $A_{LR}(2,2)$ in 
the diagonal and mixed pair--production processes. In this special case,
the 
$\tilde{\chi}^-_1\tilde{\chi}^+_1$ mode alone gives one solution
and the other contours cross at the same point which is marked by a fat dot.
In general, the cross section and asymmetry contours intersect
twice for each $(ij)$ pair combination. However,
combining the observables of the lightest pair (11) with the
second lightest pair (12) leads already to a unique 
solution [discarding accidental cases of zero measure] that can
be cross-checked again by measuring the (2,2) observables:
\begin{eqnarray}
[\cos 2\phi_L, \cos 2\phi_R] = [0.67,0.85]
\end{eqnarray}
If the three measurements could not be interpreted by a single 
$[\cos 2\phi_L, \cos2 \phi_R]$ solution, the basic set--up of the 
$2\times 2$ SUSY chargino system would have to be extended. 

In practice, the errors in the observables $m_{\tilde{\chi}^\pm_{1,2}}$
and $\cos 2\phi_{L,R}$ must be analyzed experimentally and the
migration to the fundamental SUSY parameters must be studied properly.
This is however beyond the scope of the purely theoretical
analysis in this paper.\s

\subsubsection*{4.2 Fundamental SUSY parameters}
 
{ }From the two masses $m_{\tilde{\chi}^\pm_1}$ and
$m_{\tilde{\chi}^\pm_2}$ and the mixing angles 
$\cos 2 \phi_L$ and $\cos 2 \phi_R$, the basic
SUSY parameters $\{\tan\beta, M_2,|\mu|,\cos\Phi_\mu\}$ can be 
derived  unambiguously in the following way.\\
%

\nn {\bf(i) \underline{\boldmath{$\tan\beta$:}}}
     The value of $\tan\beta$ is uniquely 
     determined in terms of two chargino masses and two mixing 
     angles
\begin{eqnarray}
\tan\beta=\sqrt{\frac{4m^2_W
                     +(m^2_{\tilde{\chi}^\pm_2}-m^2_{\tilde{\chi}^\pm_1})
                      (\cos 2\phi_R-\cos 2\phi_L)}{4m^2_W
                     -(m^2_{\tilde{\chi}^\pm_2}-m^2_{\tilde{\chi}^\pm_1})
                      (\cos 2\phi_R-\cos 2\phi_L)}}
\label{eq:tanb}
\end{eqnarray}
    For $\cos 2 \phi_R$ larger (smaller) than $\cos 2 \phi_L$  
    the value of $\tan\beta$ is larger (smaller) than unity. \\ 

\nn {\bf(ii) \underline{\boldmath{$M_2,|\mu|$:}}}
    Based on  the definition $M_2>0$, 
    the gaugino mass parameter $M_2$ and the modulus of the higgsino
    mass parameter reads as follows:
\begin{eqnarray}
 M_2&=&\frac{1}{2}
        \sqrt{2(m^2_{\tilde{\chi}^\pm_2}+m^2_{\tilde{\chi}^\pm_1}-2m^2_W)
              -(m^2_{\tilde{\chi}^\pm_2}-m^2_{\tilde{\chi}^\pm_1}) 
               (\cos 2\phi_R+\cos 2\phi_L)}\nonumber\\
 |\mu|&=&\frac{1}{2}
        \sqrt{2(m^2_{\tilde{\chi}^\pm_2}+m^2_{\tilde{\chi}^\pm_1}-2m^2_W)
              +(m^2_{\tilde{\chi}^\pm_2}-m^2_{\tilde{\chi}^\pm_1}) 
               (\cos 2\phi_R+\cos 2\phi_L)}
\label{eq:M2mu}
\end{eqnarray} 

\nn {\bf (iii) \underline{$\mbox{\boldmath{$\cos\Phi$}}_\mu$:}}
    The sign of $\mu$ in CP--invariant theories and,  
    more generally, the cosine of the phase of $\mu$ in 
    CP--noninvariant theories is determined as well by 
    the $\tilde{\chi}^\pm_1$ and $\tilde{\chi}^\pm_2$
    masses and $\cos \phi_{L,R}$:
    Using eqs.~(\ref{eq:tanb}) and (\ref{eq:M2mu}), $\cos{\Phi_\mu}$ is 
    obtained from 
\begin{eqnarray}
\cos\Phi_\mu=\frac{(m^2_{\tilde{\chi}^\pm_2}-m^2_{\tilde{\chi}^\pm_1})^2
                   -(M^2_2-|\mu|^2)^2-4m^2_W(M^2_2+|\mu|^2)
                   -4m^4_W\cos^2 2\beta}{8m^2_WM_2|\mu|\sin 2\beta}
\end{eqnarray}
%

{ } \vskip 0.3cm
As a result, the fundamental SUSY parameters $\{\tan\beta,
M_2,\mu\}$ in CP--invariant
theories, and $\{\tan\beta, M_2, |\mu|\, \cos\Phi_\mu\}$
in CP--noninvariant theories, 
can be extracted {\rm unambiguously} from the observables 
$m_{\tilde{\chi}^\pm_{1,2}}$, $\cos 2\phi_R$, and $\cos 2\phi_L$.
The final ambiguity in $\Phi_\mu \leftrightarrow 2 \pi - \Phi_\mu$ 
in CP--noninvariant theories must be resolved by measuring observables
related to the 
normal $\tilde{\chi}^-_1$ or/and $\tilde{\chi}^+_2$  polarization in 
non--diagonal (1,2) chargino--pair production \cite{R11}.

\subsection*{5. Comment on the Neutralino Sector}

Due to the large  ensemble of four neutralino states 
$[\tilde{\chi}^0_1, \tilde{\chi}^0_2, \tilde{\chi}^0_3,
\tilde{\chi}^0_4]$ in the bino--wino--higgsino sector, 
the analysis is much more complex in this case. Nevertheless,
after measuring the SU(2) gaugino mass $M_2$ and the
higgsino mass parameter $\mu$ (including the phase) in 
the chargino sector, the symmetric MSSM neutralino mass matrix
\begin{eqnarray}
{\cal M}_N = \left(\begin{array}{cccc}
      |M_1| {\rm e}^{i\Phi_1}  &      0    
    &  -m_Z s_W\cos\beta       &  m_Z s_W\sin\beta \\
                               &     M_2 
    &   m_Z c_W\cos\beta       & -m_Z c_W\sin\beta \\
    &    &     0                &    -|\mu| {\rm e}^{i\Phi_\mu} \\
    &    & &   0
           \end{array}\right)
\end{eqnarray}
involves only two unknown parameters: the modulus and the phase
of the (complex) U(1) gaugino
mass $M_1 = |M_1| e^{i\Phi_1}$.\s

Deferring the detailed analysis for a CP--noninvariant theory 
to a sequel\footnote[4]{If
$M_1$ and $M_2$ are real at the same time, what may be realized 
approximately in grand--unified scenarios, the 
subsequent analysis is modified only slightly to the extent
that $\mu^{2N+1}$ is replaced by
$|\mu|^{2N+1} \cos \Phi_\mu$ while even powers of $\mu$ are not 
altered except for the substitution
$\mu^{2N} \rightarrow |\mu|^{2N}$.} of this paper, Ref.~\cite{NEXT}, \
the analysis of CP invariant theories is much less complex.  
Since ${\cal M}^2_N$ is symmetric and positive, 
an orthogonal matrix ${\cal N}$ can be constructed that transforms 
${\cal M}^2_N$ to a positive diagonal matrix. This 
mathematical problem can be solved analytically.\s

Introducing the four--set of invariants associated with ${\cal M}^2_N$, 
%
%
%
%
\begin{eqnarray}
&& A = {\rm tr}{\cal M}^2_N
        \nonumber\\
&& { } \hskip 1cm = M_1^2+M^2_2+2\mu^2+2 m^2_Z \nonumber\\
&& B = \mbox{$\frac{1}{2}$}[({\rm tr}{\cal M}^2_N)^2-{\rm tr}{\cal M}^4_N]
        \nonumber\\ 
&& { } \hskip 1cm =(\mu^2+m^2_Z)^2+2\mu^2(M_1^2+M^2_2)+M_1^2 M^2_2 
        \nonumber\\
&& { } \hskip 1cm +2m^2_Z [c^2_W M_1^2+s^2_W M^2_2
       -\mu\sin 2\beta (c^2_W M_2+s^2_W M_1) ]\nonumber\\
&& C = \mbox{$\frac{1}{6}$}[({\rm tr}{\cal M}^2_N)^3
        -3{\rm tr}{\cal M}^2_N{\rm tr}{\cal M}^4_N
        +2{\rm tr}{\cal M}^6_N]\nonumber\\
&& { } \hskip 1cm =\mu^2[\mu^2(M_1^2+M^2_2)+m^4_Z\sin^2 2\beta
        +2M_1^2M^2_2]\nonumber\\
&& { } \hskip 1cm +m^4_Z[c^4_WM_1^2+2c^2_Ws^2_W M_1M_2+s^4_W M^2_2]
                    +2m^2_Z\mu^2(c^2_WM_1^2+s^2_WM^2_2)\nonumber\\
&& { } \hskip 1cm - 2m^2_Z\mu\sin 2\beta
              [c^2_W M_2 (\mu^2+M_1^2)+s^2_WM_1(\mu^2+M^2_2)]\nonumber\\
&& D = {\rm det}\,{\cal M}^2_N
       \nonumber\\
&& { } \hskip 1cm =\mu^4M_1^2M^2_2
       +m^4_Z\mu^2[c^4_WM_1^2+2c^2_Ws^2_WM_1M_2+s^4_W M^2_2]
                 \sin^2 2\beta\nonumber\\
&& { } \hskip 1cm -2m^2_Z\mu^3M_1M_2[c^2_WM_1+s^2_W M_2]\sin 2\beta
\label{eq:m0 determined}
\end{eqnarray}
the consistency condition 
\begin{eqnarray}
   \mc^8 - A \mc^6 + B \mc^4 -C \mc^2 + D = 0
\label{eq:mchi0}
\end{eqnarray}
must be fulfilled by the lowest of the eigenvalues
$m^2_{\tilde{\chi}^0_1}$.
Since $m_{\tilde{\chi}^0_1}$ can be measured  precisely in chargino
decays \cite{R2A}, 
\begin{eqnarray}
          \tilde{\chi}^\pm_1 \rightarrow W^\pm + \tilde{\chi}^0_1          
\end{eqnarray}
i.e. with an error of ${\cal O}$ (100 MeV), 
eq.~(\ref{eq:mchi0}) is a well--determined quadratic form that can
be solved for $M_1$ up to a 2--fold ambiguity. Moreover, it has been 
shown in Ref.~\cite{R12A} that, in fact without using further experimental 
input, the ambiguity can be removed by linearizing the consistency 
conditions in ${\cal M}_N$ instead of ${\cal M}^2_N$ in a somewhat 
involved mathematical procedure.


\subsection*{6. Conclusions}
\label{sec:conclusion}

We have analyzed how the parameters of the chargino system, 
the chargino masses $m_{\tilde{\chi}^\pm_{1,2}}$
and the size of the wino and higgsino components in the chargino 
wave--functions parameterized by two angles 
$\phi_L$ and $\phi_R$, can be extracted from pair production of 
the chargino states in $\ee$ annihilation. 
In addition to the three production cross sections, longitudinal electron 
polarization, which should be realized at  $e^+e^-$
linear colliders, gives rise to three independent LR asymmetries.
This method is independent of the chargino decay properties, i.e.
the analysis is not affected by the structure of the neutralino sector
which is very complex in extended supersymmetric theories while
the chargino sector remains isomorphic to the simple form of the
MSSM. \s 

From the chargino masses $m_{\tilde{\chi}^\pm_{1,2}}$ and the 
two mixing angles $\phi_L$ and $\phi_R$, the fundamental SUSY 
parameters $\{\tan\beta,M_2,\mu\}$ can be extracted 
in CP--invariant theories; in CP--noninvariant theories the
modulus of $\mu$ and the cosine of the phase can be determined,
leaving us with just a discrete two--fold ambiguity. The ambiguity can be
resolved however by measuring the sign of observables related to the normal 
$\tilde{\chi}^\pm_{1,2}$ polarizations.

Moreover, from the energy distribution of the final particles
in the decay of the charginos $\tilde{\chi}^\pm_1$,
the mass of the lightest neutralino  $\tilde{\chi}^0_1$ can be 
measured. This allows us to derive the parameter $M_1$ 
in CP--invariant theories so that 
the neutralino mass matrix, too, can be reconstructed
in a model-independent way. \s

In summary. The measurement of the processes
$e^+e^-\rightarrow\tilde{\chi}^-_i
\tilde{\chi}^+_j$ provides a complete analysis of the fundamental
SUSY parameters $\{\tan\beta, M_2,\mu\}$ in the chargino sector.

\subsubsection*{Acknowledgments}

This work was supported in part by the Korea Science and Engineering 
Foundation (KOSEF) through the KOSEF-DFG large collaboration project, 
Project No. 96-0702-01-01-2, and in part by the Center for Theoretical 
Physics. Thanks go to G.~Moultaka and M.~Raidal for discussions. 
We are particularly grateful to W. Bernreuther for very helpful 
discussions on disentangling CP violation in the production from the 
decay of charginos.

\bigskip \bigskip

\newpage
\mbox{ }
\vskip 5cm

\addtocounter{figure}{1}

\begin{center}
\begin{figure}[htb]
\hbox to\textwidth{\hss\epsfig{file=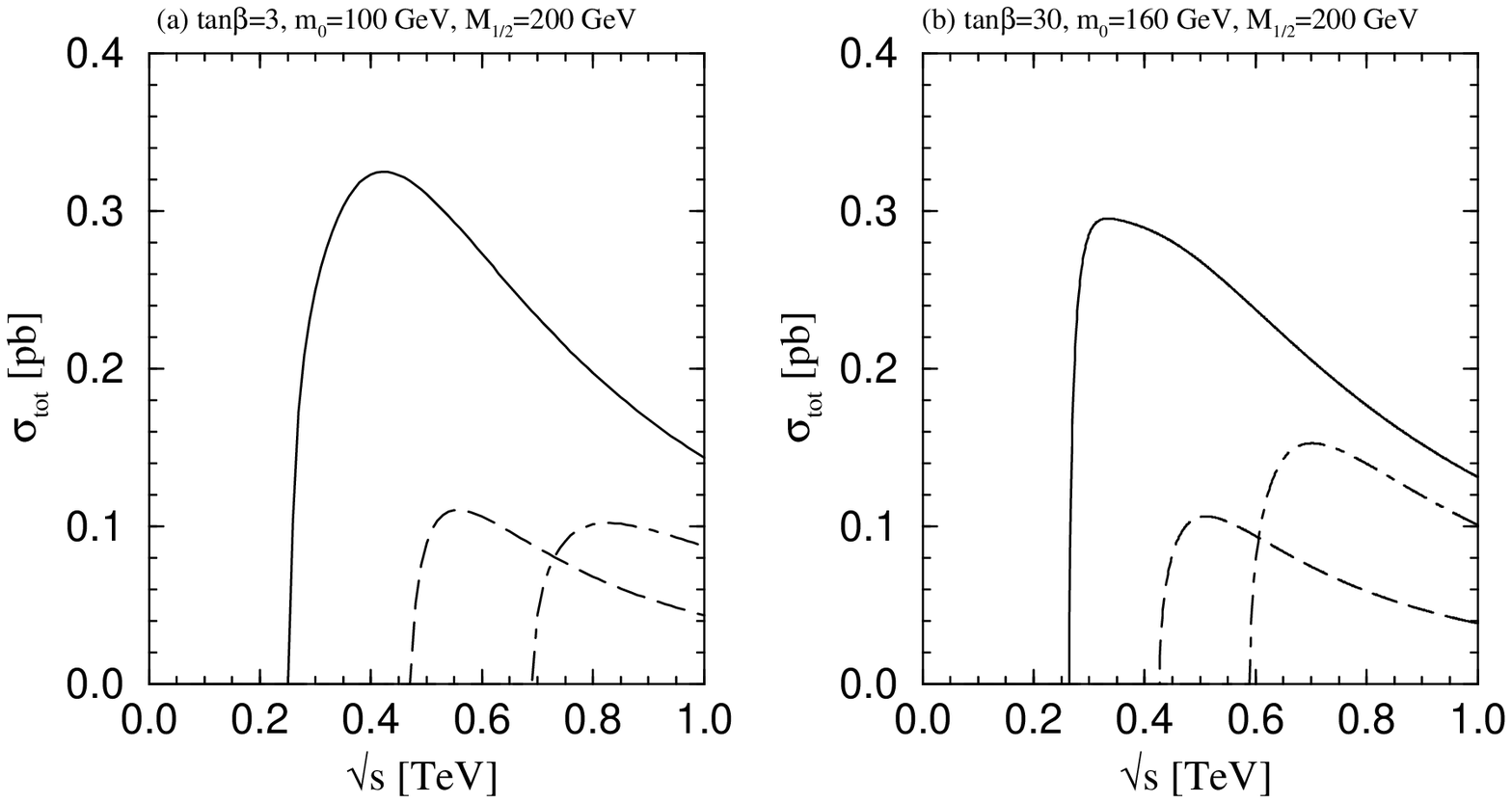,width=16cm,height=10cm}\hss}
\caption{\it The cross sections for the production of 
             charginos as a function of the c.m. energy (a) with 
             the set $[\tan\beta=3,m_0=100\, {\rm GeV},
             M_{1/2}=200\, {\rm GeV}]$ and (b) with the set  
             $[\tan\beta=30,m_0=160\,{\rm GeV},
             M_{1/2}=200\, {\rm GeV}]$: solid line for  
             $\tilde{\chi}^-_1\tilde{\chi}^+_1$ production,
             dashed line for $\tilde{\chi}^-_1\tilde{\chi}^+_2$ 
             production, and dot-dashed line for
             $\tilde{\chi}^-_2\tilde{\chi}^+_2$ production.}
\label{fig:xrs}
\end{figure}
\end{center}


\newpage
\mbox{ }
\vskip 4cm

\begin{center}
\begin{figure}[htb]
\hbox to\textwidth{\hss\epsfig{file=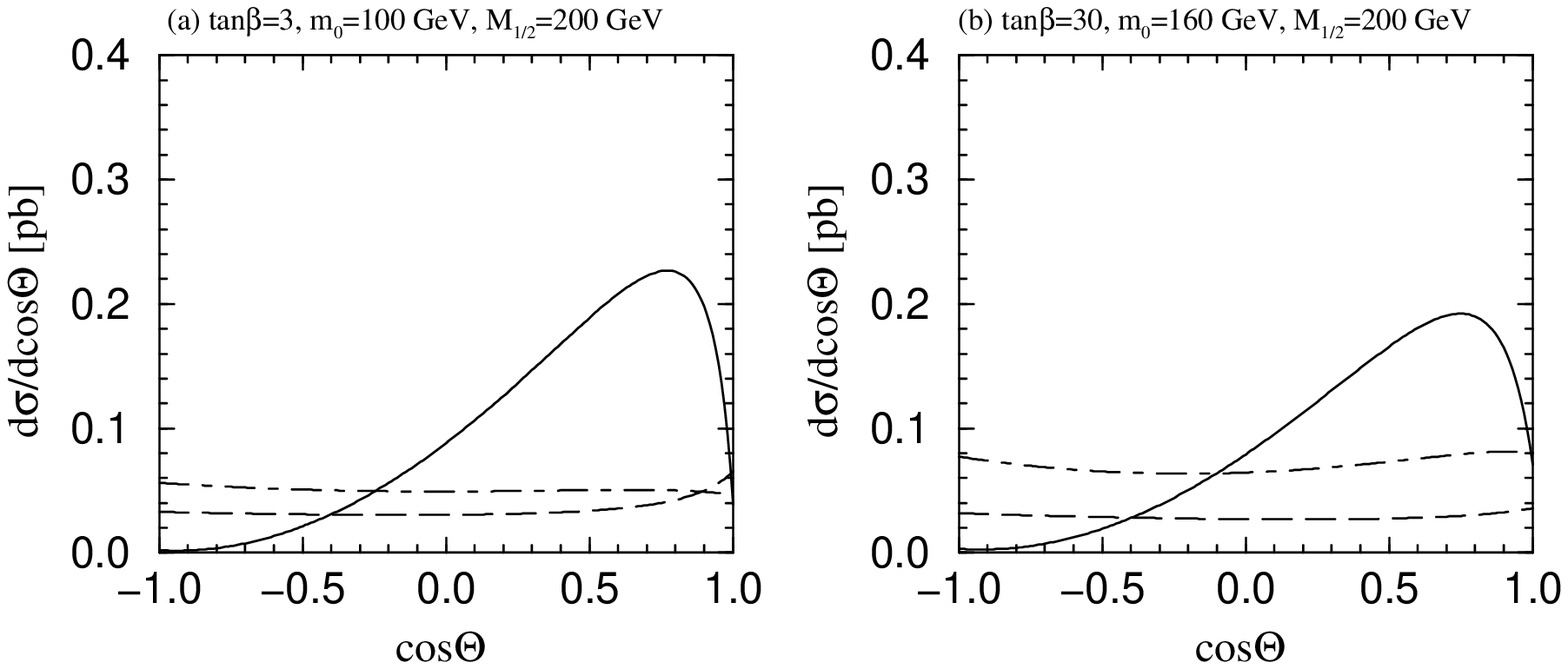,width=16cm,height=10cm}\hss}
\caption{\it The angular distributions as a function of the scattering
             angle at a c.m. energy of 800 GeV (a) with 
             the set $[\tan\beta=3,m_0=100\,{\rm GeV},
             M_{1/2}=200\,{\rm GeV}]$ and (b) with the set  
             $[\tan\beta=30,m_0=160\,{\rm GeV},
             M_{1/2}=200\,{\rm GeV}]$: solid line for 
             $\tilde{\chi}^-_1\tilde{\chi}^+_1$ production,
             dashed line for $\tilde{\chi}^-_1\tilde{\chi}^+_2$ 
             production, and dot-dashed line for
             $\tilde{\chi}^-_2\tilde{\chi}^+_2$ production.}
\label{fig:dxrs}
\end{figure}
\end{center}

\newpage
\mbox{ }
\vskip 4cm

\begin{center}
\begin{figure}[htb]
\hbox to\textwidth{\hss\epsfig{file=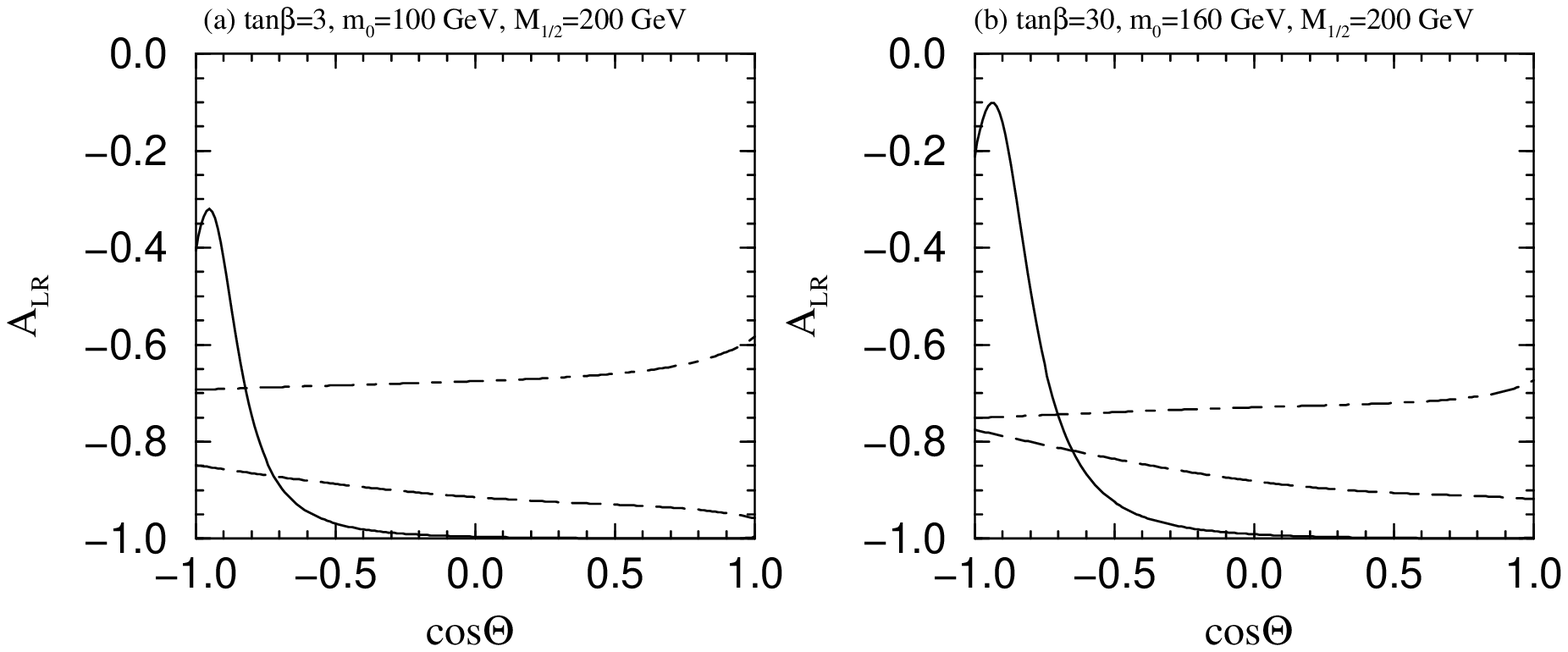,width=16cm,height=10cm}\hss}
\caption{\it The LR asymmetries as a function of the scattering angle
             at a c.m. energy of 800 GeV (a) with 
             the set $[\tan\beta=3,m_0=100\,{\rm GeV},
             M_{1/2}=200\,{\rm GeV}]$ and (b) with the set 
             $[\tan\beta=30,m_0=160\,{\rm GeV},
             M_{1/2}=200\,{\rm GeV}]$: solid line for  
             $\tilde{\chi}^-_1\tilde{\chi}^+_1$ production,
             dashed line for $\tilde{\chi}^-_1\tilde{\chi}^+_2$ 
             production, and dot-dashed line for
             $\tilde{\chi}^-_2\tilde{\chi}^+_2$ production.}
\label{fig:dlr}
\end{figure}
\end{center}

\newpage
\mbox{ }
\vskip 3cm

\begin{center}
\begin{figure}[htb]
\vskip -2cm
\hbox to\textwidth{\hss\epsfig{file=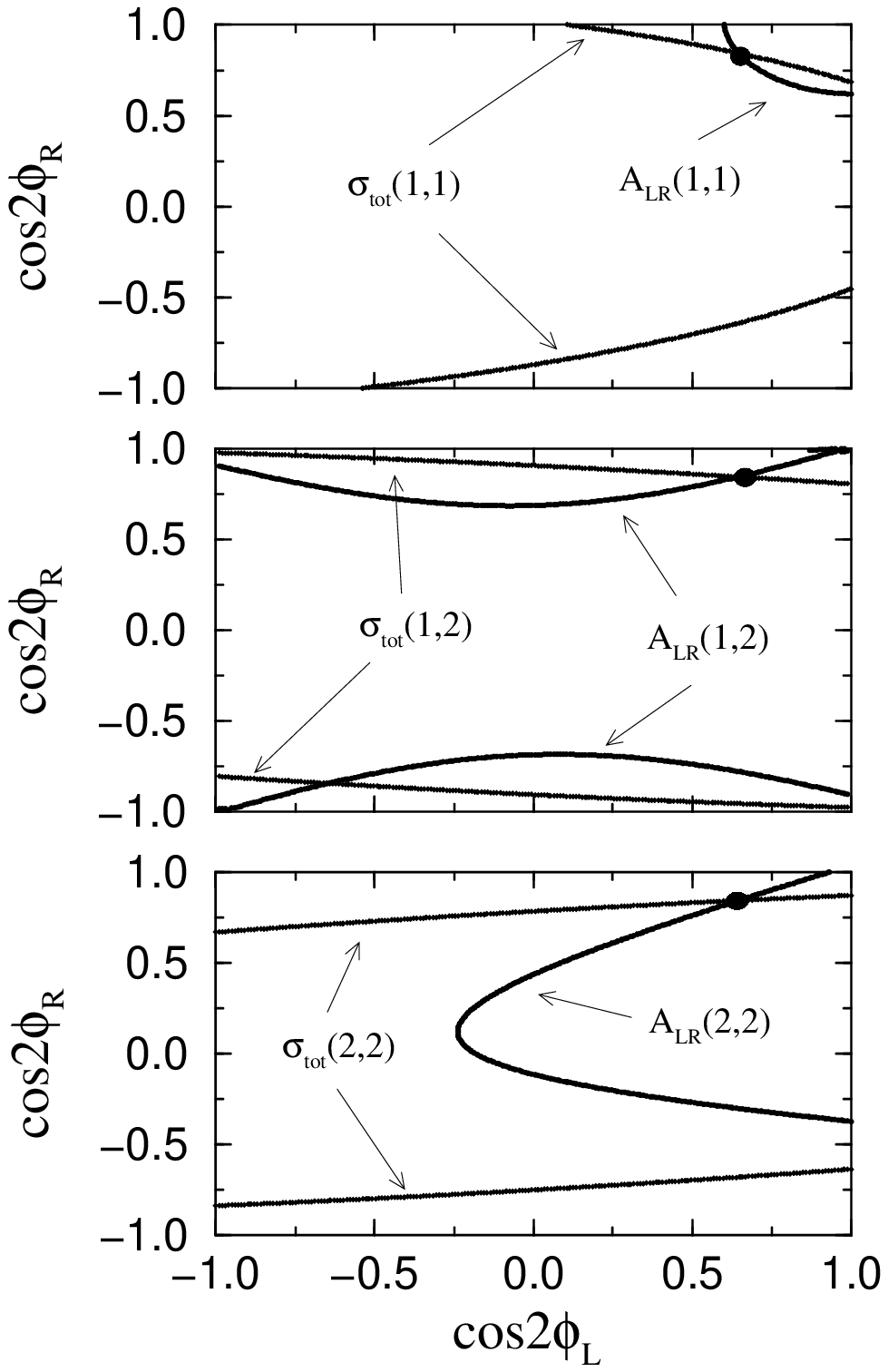,width=10cm,height=15cm}\hss}
\caption{\it Contours in the $\{\cos 2\phi_L,\cos 2\phi_R\}$ plane for 
             ``measured values" of the total cross section 
             $\sigma_{tot}(i,j)$ and the LR asymmetry $A_{LR}(i,j)$ 
             for $\tilde{\chi}^-_i\tilde{\chi}^+_j$ [$i,j=1,2$] production.
             The upper frame describes the (1,1) mode, the central frame  
             the (1,2) mode and the lower frame the (2,2) mode. 
             The fat dot in each figure marks the common crossing point 
             of the contours.}
\label{fig:cont}
\end{figure}
\end{center}
%

%

%

\end{document}